\documentclass[12pt,a4paper]{article}
\usepackage{graphicx,psfrag,bbm}
\author{P.C. Aichelburg, Ch. Lechner\\
Institut f\"ur Theoretische Physik, Universit\"at Wien\\
{\small e-mail:pcaich@doppler.thp.univie.ac.at,}\\
{\small lechner@galileo.thp.univie.ac.at}}
\title{$\sigma$-Model on de Sitter Space}
\date{} 
\newcommand{\be}{\begin{equation}}
\newcommand{\ee}{\end{equation}}
\newcommand{\bd}{\begin{displaymath}}
\newcommand{\ed}{\end{displaymath}}
\newcommand{\ba}{\begin{array}}
\newcommand{\ea}{\end{array}}
\newcommand{\beq}{\begin{eqnarray}}
\newcommand{\eeq}{\end{eqnarray}}
\newcommand{\bi}{\begin{itemize}}
\newcommand{\ei}{\end{itemize}}
\begin{document}
\begin{titlepage}
\begin{flushright}
UWThPh-1997-41\\
October 1997
\end{flushright}
\vfill
\begin{center}
{\huge $\sigma$-Model on de Sitter Space}
\vfill

{\large {\bf Peter C. Aichelburg, Christiane Lechner}}\\
\vspace{2mm}
{\large Institut f\"ur Theoretische Physik, Universit\"at Wien,\\
        Boltzmanngasse 5, A - 1090 Wien, Austria}
         e-mail: pcaich@doppler.thp.univie.ac.at,
        lechner@galileo.thp.univie.ac.at
\vfill
{\bf Abstract}
\end{center}
We discuss spherically symmetric, static solutions to the SU(2) $\sigma$-model 
on
a de Sitter background. Despite of its simplicity this model reflects many of
the features exhibited by systems of non-linear matter coupled to gravity e.g.
there exists a countable set of regular solutions with finite energy; all of
the solutions show linear instability with the number of unstable modes
increasing with energy.
\vfill
\end{titlepage}
\section{Introduction}
With the unexpected discovery of spherically symmetric static solutions to the
Einstein Yang Mills system by Bartnik and McKinnon \cite{BMK} new interest
focused on particle-like solutions for non-linear matter coupled to gravity.
Many of these systems under consideration have been shown to allow for
asymptotically flat soliton and black hole solutions 
(for a recent review on the EYM-system see \cite{EYM}). 
More recently  Brodbeck et. al. \cite{Brodbeck} investigated the 
EYM-equations including a cosmological constant and showed, that
a rich spectrum of solutions exists depending on the dimensionless parameter
in the theory.

An extremely simple model that lives on a compact 3-space was previously
considered by Bizon \cite{Bizon1}: he looked at harmonic maps from $({\mathbbm
S}^{3},G)$ to $({\mathbbm S}^{3},G)$, where $G$ denotes the metric of constant
curvature on ${\mathbbm S}^{3}$. In the present paper we modify his model 
to harmonic maps from de Sitter spacetime to ${\mathbbm S}^{3}$, i.e. we
consider a SU(2) $\sigma$-model that lives on a de Sitter spacetime with
positive cosmological constant. This model has the advantage of being more
physical while keeping the simplicity.
We look for spherically
symmetric static solutions in a static coordinate patch of this geometry.
Although the slices of constant time are ${\mathbbm S}^{3}$, 
the field equations
resulting from this model differ from those considered by Bizon. Numerical
analysis of these equations reveals the existence of a discrete sequence of
static solutions with finite energy. The energy of this sequence is bounded
from above by a limiting (singular) solution. We show that all of these
solutions are unstable under linear (time dependent) perturbations, the number
of unstable modes increasing with energy. Some care is taken to analyze the
high excitations. To check the eigenvalues of the unstable modes we compare
them to the perturbations of the limiting solution. For this we use a 
self-adjoint extension of the operator governing these perturbations.

\section{The Non-linear $\sigma$-Model}
We take $( M, g)$ to be de Sitter space (with positive cosmological
constant $\Lambda$). The non-linear $\sigma$-model under consideration is a
harmonic map 
$X:(M,g) \longrightarrow (\mathbbm{S}^3,G)$,
defined by the stationary points of the action 
$S_{\mathcal{M}} = \int d^{4}x \mathcal{L}_{\mathcal{M}}$
with the Lagrange density
\begin{equation} \label{L}
\mathcal{L}_{\mathcal{M}} = - \sqrt{ - g } \frac{f_{\pi}^{2}}{2} g^{\mu\nu}
\nabla_{\mu} X^{A}\nabla_{\nu} X^{B} G_{AB}(X).    
\end{equation}
This yields the field equations
\begin{equation} \label{equ}
g^{\mu\nu}(\nabla_{\mu}\nabla_{\nu}X^{A} + {\tilde \Gamma}^{A}_{BC}
\nabla_{\mu}X^{B}\nabla_{\nu}X^{C}) = 0,
\end{equation}
where ${\tilde \Gamma}^{A}_{BC}$ denote the Christoffel symbols with respect to
the metric $G_{AB}$ of the target manifold. We choose coordinates
$(X^{A})=(f,\Theta,\Phi)$, so that $G$ assumes the form $G_{AB} = \textrm{diag}
(1, \sin^{2}f, \sin^{2}f \sin^{2}\Theta)$.
%
%
%
It is well known that part of the de Sitter space can be covered by
a static frame i.e. coordinates $(t,r,\vartheta, \varphi)$ in which the metric
becomes
\begin{equation} \label{g}
ds^{2} = -\left( 1 - \frac{r^{2}}{\alpha^{2}}\right) dt^{2} 
+ \left( 1 - \frac{r^{2}}{\alpha^{2}}\right)^{-1} dr^{2} + r^{2}d\vartheta^{2}
+ r^{2}\sin^{2} \vartheta d\varphi^{2},
\end{equation}
with $0 \le r < \alpha$ and $\alpha^{2} = 3/\Lambda$, $\Lambda$ being the
cosmological constant.
The singularity  at $r = \alpha$ signals the breakdown of the static
coordinates.
We seek for regular solutions of Eqs. (\ref{equ}) in the domain, 
where the metric takes
the static form (\ref{g}) and require the mapping to be static, i.e. 
$\partial_{t} X^{A} = 0$, and spherically symmetric. 
The latter is guaranteed by the
Hedgehog-Ansatz:
\begin{equation} \label{U}
f(x^{\mu})=f(r), \Theta(x^{\mu})= \vartheta, \Phi(x^{\mu}) = \varphi.
\end{equation}
In order to map the pole $r=0$ to a single point of ${\mathbbm S}^{3}$, 
we require the
function $f(r)$ to obey the boundary condition $f(0)=0$.
Inserting this Ansatz into Eqs.(\ref{equ}) yields the non-linear ordinary
differential equation
\begin{equation} \label{f}
\left( r^{2} \left( 1 - \frac {r^{2}}{\alpha^{2}}\right) f' \right)' - \sin 2f
= 0.
\end{equation}
The energy associated with an excitation of the form (\ref{U}) is given by
\begin{equation} \label{E}
E = 4 \pi f^{2}_{\pi} \int_{0}^{\alpha} r^{2}dr 
        \lbrace (1 - \frac {r^{2}}{\alpha^{2}})
f'^{2} + \frac{2}{r^{2}} \sin^{2} f \rbrace 
\end{equation}
%
where a factor of $2$ has been included, because the $r$-integration runs twice
from $0$ to $\alpha$.\\
%
%
%
Note, that the energy functional $E[f]$ (\ref{E}) may be interpreted again as 
a spherically
symmetric harmonic map, this time from $({\mathbbm S}^{3},(1 -\frac {r^{2}}
{\alpha^{2}})^{2}G)$ 
to $({\mathbbm S}^{3},G)$. 

To our knowledge the only analytic solutions to (\ref{f}) are the
trivial solution $f \equiv 0 $ (mod $\pi$) 
 with energy $E_{0} = 0$
and the  solution $f \equiv \pi/2$, which will be called singular, since it 
fails to obey the boundary condition at
$r = 0$ (see section \ref{singsol}).

\section{Numerical Results}
In the following we introduce a new independent variable $x$ defined by 
$ r = \alpha \sin \chi$
and $ x = - \ln {\tan {\chi/2}}$. The coordinate $x$ tends from $x = 0$ 
($r = \alpha$) to
$x = \infty$ ($r = 0$). We also define a new dependent variable $h$  by
$h = f - \pi/2$.
In these new variables equation (\ref{f}) becomes
\begin{equation} \label{h}
h'' - \left( \tanh x - \frac {1}{\sinh x \cosh x} \right) h' + \sin {2h} = 0,
\end{equation}
where prime now denotes differentiation with respect to $x$.

Equation (\ref{h}) is similar to the one studied by
P.Bizon \cite{Bizon1}.
Ours differs from his by the term
$(\sinh x \cosh x)^{-1}$ which causes a singular point of the differential
equation at $x = 0$  (corresponding to the breakdown of the static frame at
$r= \alpha$). For the solutions to be regular at $x = 0$, $h'(0)$ has to vanish.
The other singular point is at $x = \infty$ where 
$h(\infty) = \pm \pi/2$ (mod$\pi$).
The formal power series expansions of $h(x)$ around these singular points give:
\be
\begin{array}{rcll}
h(x) & = & b - 1/4 (\sin {2b}) x^{2} + O(x^{4}) & 
\quad \textrm{for } x \simeq 0
\qquad\textrm {and}\\
\pm h(x)& = & \pi/2 - 2c e^{-x} + O( e^{-3x})&
\quad\textrm{for } x\rightarrow \infty .
\end{array}
\ee
One can show that near $x = 0$ as well as $x=\infty$, 
there exists a family of solutions, each
of them depending only on a single parameter, $b \equiv h(0)$ and $c$ 
respectively.
Since with $h(x)$ also $\tilde h(x) = n\pi \pm h(x)$ 
is solution to equation (\ref{h}), we restrict the parameter $b$ to the range 
$0 < b <\pi/2 $.

We integrated equation (\ref{h}) numerically, starting at $x = 0$, trying to
find values for $b$, such that $h(x)$ shows the correct asymptotic behavior
for large $x$.
With the values found for $b$ we used the two-point shooting and matching 
method to determine the parameter $c$. Both numerical routines, d02cbf and 
d02agf,
were provided by the NAG-Library \cite{NAG}.
%
%
%
%
The numerical analysis suggests the existence of a countable family of
globally regular solutions of Eq.(\ref{h}) which may be   
labeled 
by their number of zeros, and indeed a rigorous 
proof of existence can be given along the
lines in \cite{Bizon1}. 

\begin{figure}[h]
\begin{center}
\begin{psfrags}
 \psfrag{-1.0}[]{$-1.0$}
 \psfrag{1.0}[]{$1.0$}
 \psfrag{0.0}[]{$0.0$}
 \psfrag{5.0}[]{$5.0$}
 \psfrag{15.0}[]{$15.0$}
 \psfrag{10.0}[]{$10.0$}
 \psfrag{20.0}[]{$20.0$}
 \psfrag{x}[]{$x$}
 \psfrag{-pi/2}[]{$-\frac{\pi}{2}$}
 \psfrag{h(x)}[][]{$h(x)$}
\includegraphics[width=4in]{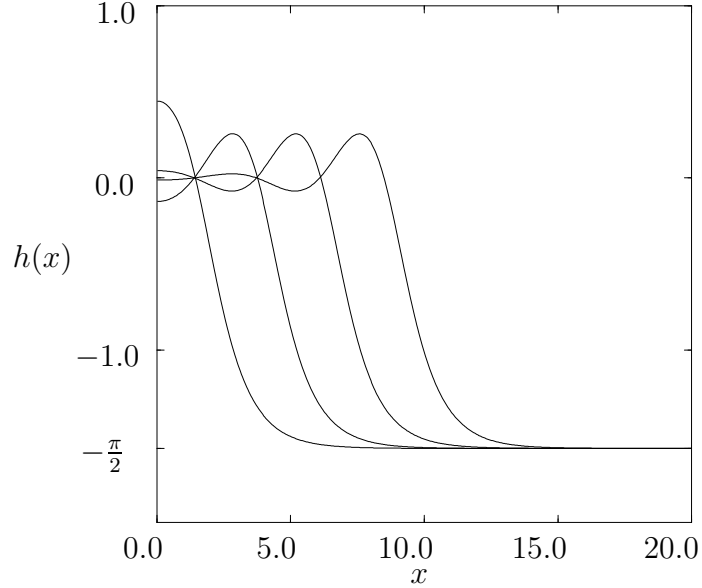} 
\end{psfrags}
\end{center}
\caption{The first four solutions $h_{n}(x)$ (The sign of $b_{n}$ has been
chosen such that $h_{n}(\infty)= -\frac{\pi}{2})$.}\label{hnx}
\end{figure}

Figure \ref{hnx} shows the first four solutions
$h_{n}(x)$, where the sign of $b_{n}$ has been chosen such that
$h_{n}(\infty) = - \pi/2$.
The numerical results $b_{n}$ and $c_{n}$ 
for the first four solutions are displayed in Table \ref{table1}.

\begin{table}[htbp]
\begin{center}                                                                                                             
\begin{tabular}{|c|c|c|c|c|c|}
\hline
$n$ & $b_n$ & $c_n$ & $b_{n}^{2}/b_{n+1}^{2}$ & $c_{n+1}/c_{n}$ & $E_{n}/4\pi
f_{\pi}^{2}\alpha$ \\
\hline
1 & 0.4458536 & 5.0270678 & 10.636094 & 10.831606 & 1.8304231 \\
2 & 0.1367103 & 54.451223 & 10.734453 & 10.747914 & 1.9842302 \\
3 & 0.0417264 & 585.23706 & 10.747677 & 10.748889 & 1.9985328 \\
4 & 0.0127278 & 6290.6484 & 10.748955 & 10.749068 & 1.9998635 \\
`$\infty$'&   &           & 10.749087 & 10.749087 & 2         \\
\hline
\end{tabular}
\end{center}
\caption{Numerical values of the parameters $b_{n}$, $c_{n}$, the ratios
$b_{n}^{2}/b_{n+1}^{2}$,$c_{n+1}/c_{n}$ and the energy $E_{n}$ of the first
four excitations.}\label{table1} 
\end{table} 
%
%
%
%
%
%
%
\section{Large Excitations}
In what follows we give arguments for understanding the numerical value of the
ratios formed from the integration constants for consecutive values of $n$ as
listed in Table \ref{table1}. The line of reasoning, which is based on
approximating Eq. (\ref{h}) for large excitations by linear respectively
translation
invariant differential equations, is the same as in \cite{Bizon1},
but for completeness we will repeat it here.

For large n one can distinguish essentially three regions for equation
(\ref{h}):

{\em Region I}, for $0 \leq x \leq 4$, where numerical analysis tells us, that 
$h$ is small, so that Eq.
(\ref{h}) is well approximated by
\begin{equation} \label{I}
 h'' - \left( \tanh x - \frac {1}{\sinh x \cosh x} \right)h' + 2h = 0.
\end{equation}
Since equation (\ref{I}) is linear $h_{n}(x) \simeq b_{n}g(x)$ for large n,
where $g(x)$ solves equation (\ref{I}) with initial conditions $g(0) = 1$ and
$g'(0) = 0$.

In {\em region II}, $4 \leq x \leq 2n$, where $h$ is still small, 
one can approximate
equation (\ref{h}) for large $n$ by
\begin{equation} \label{II} 
h'' - h' + 2h = 0,
\end{equation}
which is linear as well as translation invariant. Eq. (\ref{II}) is solved by
$h_{n} \simeq b_{n} g(x)$ with $g(x) \propto e^{x/2} 
\sin {(\sqrt{7} x/2  + \delta)}$.

Finally  in {\em region III}, for $x \geq 2n$, equation (\ref{h}) becomes
\begin{equation} \label{III} 
h'' - h' + \sin {2h} = 0,
\end{equation}
and is solved asymptotically by $\pm h_{n}(x) = \pi /2 - 2 c_{n} e^{-x}$.
%
%

Now, by using translation invariance of Eqs. (\ref{II}) and (\ref{III}) as well
as the asymptotic behavior of $h_{n}$, one obtains in region II and III for two
consecutive solutions:
%
\begin{equation} \label{a}
h_{n+1}(x)\simeq h_{n}(x - \ln {c_{n+1}/c_{n}}),
\end{equation}
if we choose $h_{n+1}$ and $h_{n}$ to have the same asymptotic behavior.
(which is possible by choice of sign).
%
In the regions I and II, due to the linearity of Eqs. (\ref{I}) and 
(\ref{II}) and $h_{n}(0)=0$,
$h_{n+1}$ and $h_{n}$ are connected by
\be
h_{n+1} = \frac{b_{n+1}}{b_{n}}h_{n}.
\ee
Using the solution to Eq. (\ref{II}) and the fact, that,
in region II, $h_{n+1}$ has one zero more than $h_{n}$, one determines the
parameter values of consecutive solutions to behave like
\begin{equation} \label{lim}
b_{n}^{2}/b_{n+1}^{2} \simeq c_{n+1}/c_{n} \simeq e^{2\pi /\sqrt{7}}
\end{equation}
for large $n$.
 
%
%
%
%
%
%
\section{Stability Analysis}
We now discuss linear stability of the solutions $h_{n}$ under 
time dependent
spherically symmetric perturbations of $U$, i.e. we use the Ansatz:
\begin{equation} \label{A}
h(r,t) = h_{n}(r) + \delta h(r,t)
\end{equation}
where $h_{n}(r)$ is solution to the static field equation (\ref{h}) and
$\delta h(r,t)$ is considered to be small, so that we retain only terms linear
in $\delta h$. The full time dependent field equation is:
\begin{equation} \label{z}
\left( r^{2} \left( 1 - \frac {r^{2}}{\alpha^{2}} \right) h' \right) ' + 
\sin {2h} = \frac {r^{2}}{\left( 1 - \frac {r^{2}}{\alpha^{2}} \right)} \ddot
h.
\end{equation}  
Inserting (\ref{A}) into (\ref{z}) and linearizing in $\delta h$ gives:
\begin{equation} \label{st}
\left( r^{2} \left( 1 - \frac {r^{2}}{\alpha^{2}} \right) \delta h' \right) ' + 
2 \cos (2h_{n}) \delta h = \frac {r^{2}}{\left( 1 - \frac {r^{2}}
{\alpha^{2}} \right)} \delta \ddot h.
\end{equation}
We set 
\begin{equation} \label{d}
\delta h(r,t) = \xi(r) e^{-i \sigma t},
\end{equation}
which leaves us with
\begin{equation}\label{b}
\left( r^{2} \left( 1 - \frac {r^{2}}{\alpha^{2}} \right) \xi' \right) ' + 
2 \cos (2h_{n}) \xi = \frac { - \sigma^{2}r^{2}}{\left( 1 - \frac {r^{2}}
{\alpha^{2}} \right)} \xi.
\end{equation}
If this eigenvalue problem has any modes with $\sigma^{2} < 0$ , $\delta h$ 
would grow exponentially in time and our solutions would be linearly
unstable.

The energy of the time-dependent field is given by 
\begin{displaymath}
E = 4 \pi f^{2}_{\pi} 
          \int_{0}^{\alpha} r^{2} dr \left( \frac {r^{2}}
           {\left( 1 - \frac {r^{2}}{\alpha^{2}} \right)} \dot h^{2} + 
          \left( 1 - \frac {r^{2}}{\alpha^{2}} \right) h'^{2} + 
          \frac{2}{r^{2}}\cos^{2}{h} \right)
\end{displaymath}
Regular perturbations have to satisfy the boundary conditions $\xi (\alpha) =
0$ and $\xi (0) = 0$.
%
%
%
%
Using the coordinate $x$ Eq. (\ref{b}) reads
\be \label{xix}
\sinh^{2}(x)\xi'' + \tanh (x)(1-\sinh^{2}(x))\xi' + 2\cos (2h_{n})\sinh^{2}(x)\xi
        = - \lambda \xi,
\ee
where $\lambda = \sigma^{2} \alpha^{2}$.

The asymptotic behavior of $\xi$ is:
\be
\begin{array}{rcll}
 \xi & \sim & x^{\sqrt{-\lambda}} & \quad \textrm {for $x \sim 0$ and}   \\
 \xi & \sim & e^{-x}   & \quad \textrm {for $x \rightarrow \infty$.}
\end{array}
\ee
Using routine d02kdf of the Fortran NAG-Library \cite{NAG} we find numerically 
(up to $n=3$)
that the $n$-th "excitation" has $n$ negative eigenvalues, i.e. $n$ unstable
modes. The numerical values of the first three negative eigenvalues of Eq.
(\ref{xix}) for the first six excitations 
are displayed in Table
\ref{table2}. 
\begin{table}[htbp]
\begin{center}                                                                                                             
\begin{tabular}{|c|c|c|c|}
\hline
$n$ & $\lambda^{n}_{1}$ & $\lambda^{n}_{2}$ & $\lambda^{n}_{3}$  \\
\hline
1 & -2.111967 &  &  \\
2 & -2.065756 & -183.16864 &  \\
3 & -2.061672 & -180.59393 & -21007.87 \\
4 & -2.061294 & -180.3407 &  -20715.26   \\
5 & -2.061270 & -180.3169 &  -20686.48  \\
6 & -2.061256 & -180.3148 &  -20683.78  \\
\hline
\end{tabular}
\end{center}
\caption{Numerical values of the first three negative eigenvalues}\label{table2}
\end{table}
\section{The Limiting Solution}\label{singsol}
From Eq.(\ref{lim}) we conclude that the amplitudes in region I and II decrease
with $n$ as $e^{-n\pi/\sqrt{7}}$. At the same time the region of oscillations
extends to larger and larger values of $x$. Thus for $n \to \infty$ the
solution $h_{n}(x)$ tends point wise to zero for any finite $x$.

The limiting solution $h_{\infty}(x)=0$ does not satisfy the required boundary
conditions at $x = \infty$, and as a consequence its energy density is
singular. Nevertheless the total energy (\ref{E}) is finite, 
$E_{\infty} = 8\pi
f_{\pi}^{2} \alpha$, which can be shown to be the upper bound for the energy of
all regular solutions $h_{n}$:\\
Expressing the static Eq. (\ref{f}) in terms of $h$, multiplying by $h$ leads
after a partial integration to the relation
\be
\int\limits_{0}^{\alpha} dr \ r^{2} (1 - \frac{r^{2}}{\alpha^{2}}) {h'}^{2} = 
\int\limits_{0}^{\alpha}dr \ h \sin 2h.
\ee
Thus for regular solutions the energy functional $E[h]$ can be written as
\be\label{En.expr}
E[h] = 4 \pi f_{\pi}^{2} \int\limits_{0}^{\alpha} dr (h \sin 2h + 2 \cos^{2}
h).
\ee
Since the regular solutions satisfy $|h| \le \pi/2$, the integrand of
(\ref{En.expr}) has a maximum for $h=0$.

The reason for our interest in the singular solution $h(x) =0$ is, that
the eigenvalues of the linear perturbations about regular static solutions seem
to converge to some value $\lambda_{k}^{\infty}$ for fixed $k$ as $n \to
\infty$. (See Table \ref{table2}). So the question is, can we find the limiting
values by calculating perturbations of the singular solution? This would yield a
check of the numerics and moreover tell us, that we really have obtained all
eigenvalues in the stability analysis.

Let us first have a look at Eq. (\ref{xix}) in the limit $n \to \infty$, i.e.
$h_{\infty} \equiv 0$. Introducing a new variable $w = - \sinh^{2} x$ and
setting $\xi(w) = (-w)^{\sqrt{|\lambda|}/2}y(w)$ leaves us with a
hyper-geometric differential equation
\begin{eqnarray}\label{yw}
w(1-w)y''(w) & + &\left( \sqrt {|\lambda|} + 1 -w( \sqrt {|\lambda|} + 
\frac {1}{2}) \right)y'(w) + \nonumber\\
  & & + \left(  \frac {\sqrt {|\lambda|}}{4} - \frac{ |\lambda|}{4} -
\frac{1}{2} \right)y(w) = 0,
\end{eqnarray}
which is solved by the hyper-geometric Function 
$F(a,b,c;w) = \sum_{n=0}^{\infty} \frac {(a)_{n}(b)_{n}}{n!(c)_{n}} w^{n}$,
where $(a)_{n} = a(a+1)\cdots(a+n-1)$; $(a)_{0} = 1$ etc. and
\bd
a= -1/4 + \sqrt{|\lambda|}/2 + i\sqrt{7}/4,
\ed
\bd
b=-1/4 + \sqrt{|\lambda|}/2 - i\sqrt{7}/4,
\ed
\bd
c=1+\sqrt{|\lambda|}.
\ed
Analytic extension of $F(a,b,c;w)$ for $|w| > 1$ gives
\begin{eqnarray}\label{as}
\xi(w) & =& (-w)^{1/4 - i\sqrt{7}/4} P(\lambda)
            F(a,1-c+a,1-b+a; 1/w)  + \nonumber\\
        & &  + (-w)^{1/4 + i\sqrt{7}/4} {\bar {P}}(\lambda)
            F(b,1-c+b,1-a+b;1/w), 
\end{eqnarray}
where ${\bar {{}}}$ denotes complex conjugation and 
\bd
P(\lambda) = \frac{\Gamma (c) \Gamma(b-a)}{\Gamma(b) \Gamma(c-a)}.
\ed
%
%

In order to get a spectrum of eigenvalues, that may serve as a limit to
the spectra of regular perturbations, we need to specify boundary conditions
at $ w = -\infty$ ($ r = 0$). This is not straightforward, since 
the solution $h_{\infty} \equiv 0$ is already singular.  
In what follows we make use of an
idea which is due to P. Bizon \cite{biz2}:

We first note, that the differential operator appearing in equation (\ref{xix}),
rewritten in the coordinate $ w = - \sinh ^{2}x$,
%
\be \label{tau}
\tau_{n} := 4 w (1 - w)^{3/2}
\frac{d}{dw}\left( \frac {w}{(1 - w)^{1/2}}
\frac{d}{dw}
\right) - 2 w \cos (2h_{n}) 
\ee
is formally self-adjoint with respect to the scalar product
%
\be\label{scalpr}
\langle \xi,\eta \rangle = \int_{-\infty}^{0}\frac{dw}
      {(-w)(1 - w)^{3/2}}{\bar \xi}\eta,
\ee
i.e.
%
%
\be\label{selfad}
\langle \tau_{n} \xi,\eta \rangle = \langle \xi, \tau_{n} \eta \rangle 
      + \left. \left( \frac {4(-w)}{(1 - w)^{1/2}} 
      ( ({\bar \xi})' \eta - \eta' {\bar \xi} )
      \right) \right|_{- \infty}^{0}.        
\ee      
The boundary terms vanish for (regular) perturbations of regular solutions
$h_{n}$. It is this property that we would like to impose on the singular
perturbations. For solutions of the eigenvalue problem (\ref{b}) this
requirement yields the condition, that the ratio of coefficients in (\ref{as})
$\frac{\bar {P}}{P} = \textrm{e}^{- 2 i \mbox{\footnotesize arg}(P)}$
has to be independent of $\lambda$. In fact, admitting solutions that behave
for $w \to -\infty$ as in (\ref{as}) with  $\frac{\bar {P}}{P}$ fixed,
corresponds to a self-adjoint extension of the operator $\tau_{\infty}$, (see
Appendix).

We are now free to choose one of the eigenvalues,
say
$\lambda_{1}^{\infty} = -2.06126$. 
The spectrum of $\tau_{\infty}$ then 
is determined
by $\textrm{e}^{2 i \mbox{\footnotesize{arg}}(P(\lambda_{k}))} =
 \textrm{e}^{2 i \mbox{\footnotesize{arg}}(P(\lambda_{1}))}$,
or equivalently
\be\label{argP}
\textrm {arg}(P(\lambda_{k})) = \textrm{arg}(P(\lambda_{1})) + (k - 1)\pi.
\ee
We obtain e.g. 
 $\lambda_{2}^{\infty} = -180.315$ and
$\lambda_{3}^{\infty}= - 20683$ in good agreement with Table \ref{table2}.

%
%
%
\section{Outlook}
In this paper we have analyzed spherically symmetric solutions to the non
linear sigma model on de Sitter space with a positive cosmological constant.
These solutions are regular everywhere in a static coordinate patch of this
spacetime i.e. up to the cosmological horizon. Despite the simplicity of the
model it shares many of the properties of matter models coupled to gravity and
may therefore help to understand why such excitations exist. Furthermore we
expect that when the back reaction is taken into account, at least for weak
coupling, these solutions will persist.

The authors would like to thank P. Bizon, S. Husa and H. Narnhofer for 
helpful discussions, moreover we
wish to thank S. Husa for his support regarding the numerical
computations. 
\appendix
\section{Appendix}
The perturbation equations (\ref{b}) can be transformed into a one-dimensional
Schr\"odinger equation by introducing new dependent and independent variables:
\be
\xi = \frac{\phi}{r} \quad \textrm{and} \quad \rho = \frac{1}{2}\ln
\frac{\alpha + r}{\alpha -r}
\ee
Eqs. (\ref{b}) then become
\be
-\frac{d^{2}}{d\rho^{2}}\phi - 2\frac{(\cos(2h_{n}) +
\tanh^{2}\rho)}{\sinh^{2}\rho}\phi = \lambda \phi.
\ee
For small $\rho$ the potential behaves as $\frac{2}{\rho^{2}}$ for regular
perturbations and as $-\frac{2}{\rho^{2}}$ for the singular case $h_{\infty}
\equiv 0$.

In order to determine domains (if possible) on which the operators are
self-adjoint, it is sufficient to consider the operators $\tau_{0} =
-\frac{d^{2}}{d\rho^{2}} + \frac{2}{\rho^{2}}$ for the regular case,
$\tau_{0,\infty} =
-\frac{d^{2}}{d\rho^{2}} - \frac{2}{\rho^{2}}$ for the singular case
respectively. These operators have been studied by H. Narnhofer
\cite{Narnhofer}:

It turns out that the operator $\tau_{0}$ is self-adjoint on the domain 
\be
\mathcal{D}(\tau_{0}) = \lbrace \phi, \phi' \textrm{abs. cont.}, \phi'' \in 
\mathcal{H}, \phi(0)= \phi'(0) = 0 \rbrace,
\ee
where $\mathcal{H} = L^{2}(d\rho, (0,\infty))$.

$\tau_{0,\infty}$ is hermitian but not self-adjoint on the above domain, since
$\textrm{dim(Ker}(\tau_{0,\infty}^{\ast} \mp i)) = n_{\pm} = 1$.
Solutions of $(\tau_{0,\infty}^{\ast} \mp i)\phi_{\pm} = 0$ are
$\phi_{+} \propto \rho^{1/2}(J_{i\sqrt{7}/2}(\rho \textrm{e}^{i\pi/4}) - 
\textrm{e}^{-\sqrt{7}\pi/2}J_{-i\sqrt{7}/2}(\rho \textrm{e}^{i\pi/4}))$, 
$\phi_{-} \propto {\bar \phi}_{+}$ respectively.

By adding linear combinations of $\phi_{\pm}$ of the form
\be\label{ext}
\phi_{+} + \textrm{e}^{2i\gamma}\phi_{-}
\ee
to the domain $\mathcal{D} (\tau_{0})$ one obtains a one paramter family of
self-adjoint extensions $\tau_{\gamma}$ of the operator $\tau_{0,\infty}$.
Solutions (\ref{as}) of the eigenvalue equation belong to $\mathcal{D}
(\tau_{\gamma})$, if one
chooses $\frac{\bar {P}}{P} =  2^{-i\sqrt{7}/2} \Gamma(1 - i\sqrt{7}/2)
\sin(\gamma - \frac{i\sqrt{7}\pi}{8})/\Gamma(1 + i\sqrt{7}/2)
\sin(\gamma + \frac{i\sqrt{7}\pi}{8})$.


\end{document}